\documentclass[12pt]{article}
\usepackage{amsfonts}
\usepackage{amsmath}
\usepackage{epsfig}
\usepackage{latexsym}
\usepackage{amssymb}
\usepackage{graphicx}

\def\ket #1{\vert #1\rangle}

\def\ketbra #1#2{\vert #1\rangle \langle #2\vert}
\newcounter{defin}
\newcounter{lemma}
\newcounter{theorem}
\newcounter{proposition}
\newcounter{example}
\newenvironment{lemma}{\par\refstepcounter{lemma}     \textbf{Lemma \thelemma.} }{\rm\par}
\newenvironment{theorem}{\par\refstepcounter{theorem}     \textbf{Theorem \thetheorem.}\ }{\rm\par}
\newenvironment{proposition}{\par\refstepcounter{proposition}     \textbf{Proposition \theproposition.}\ }{\rm\par}

\begin{document}

\title{The information capacity of entanglement-assisted continuous variable
measurement}
\author{Holevo~A.\,S., Kuznetsova~A.\,A. \\
Steklov Mathematical Institute, RAS, Moscow, Russia}
\date{}
\maketitle

\begin{abstract}
The present paper is devoted to investigation of the entropy reduction and
entanglement-assisted classical capacity (information gain) of continuous
variable quantum measurements. These quantities are computed explicitly for
multimode Gaussian measurement channels. For this we establish a fundamental
property of the entropy reduction of a measurement: under a restriction on
the second moments of the input state it is maximized by a Gaussian state
(providing an analytical expression for the maximum). In the case of one
mode, the gain of entanglement assistance is investigated in detail.
\end{abstract}

\section{Introduction}

\label{s1}

Continuous variable (CV) systems constitute one of the prospective platforms
for implementation of quantum communication and computation protocols \cite%
{CONT}, \cite{WEEDBROOK}, \cite{sera}. During the past few years, an
important chapter of quantum information science -- quantum Shannon theory
-- is being developed for CV systems, which requires mathematical tools of
infinite-dimensional Hilbert space. The theory of various channel capacities
and related entropic quantities was elaborated, in particular, for bosonic
Gaussian channels (see e.g. \cite{QSCI} and references therein).

The notion of quantum channel presupposed quantumness of both the input and
output systems, making necessary a separate treatment of \emph{\ quantum
observables}, which do not allow a simple reduction to quantum channels in
the CV case (contrary to the discrete finite-dimensional case). In
particular, this fully applies to quantum bosonic Gaussian observables. Thus
we are led to the study of \emph{quantum measurement channels}, which map CV
quantum input into CV classical output, and to computation of their
information-processing and entropic characteristics.

An important quantity characterizing information-processing performance of
the quantum measurement channel is its classical capacity \cite{da}, \cite%
{Oreshkov}, \cite{h5}. The computation of the classical capacity for
multi-mode quantum Gaussian measurement channels, based on the progress in
the solution of the \emph{quantum Gaussian optimizer conjecture} \cite{ghg},
\cite{ghm}, was recently developed in \cite{acc} under the assumption of
global gauge symmetry (\textquotedblleft phase
insensitivity\textquotedblright ), and in \cite{acc-noJ} under certain more
general \textquotedblleft threshold condition\textquotedblright .

In the present paper we study another important characteristic of a CV
measurement channel -- the \emph{entropy reduction} \cite{Gro}, \cite{Oz},
which is strongly related to its quantum mutual information \cite{winmas},
\cite{Sh-ERQM} and to the \emph{entanglement-assisted classical capacity}
\cite{h5}, \cite{H1}, \cite{H2}. In finite dimensions related notions were
studied by a number of authors under the names \emph{purification capacity,
measurement strength, information gain} of the measurement (see \cite{BRW}
where one can find also a detailed survey of the subject and further
references). 
Thus the entropy reduction and the entanglement-assisted classical capacity
of a quantum measurement are of considerable interest from various points of
view in quantum Shannon theory.

By using results previously obtained in \cite{H1}, \cite{H2}, we derive here
computable expressions for the entropy reduction and entanglement-assisted
capacity of multimode Gaussian measurement channels. We prove that under a
restriction on second moments, the entropy reduction of Gaussian observable
is maximized by a Gaussian state, explicitly giving the value of the
maximum. This fundamental property of the entropy reduction is parallel to a
similar property of quantum mutual information for quantum Gaussian
channels, however the proof is somewhat more intricate due to absence of the
Schmidt decomposition and symmetry between parts of a composite hybrid
(classical-quantum) system. As an application we consider in detail the case
of one mode and study the gain of the entanglement-assisted vs unassisted
classical capacities of the measurement channel. Our findings give another
evidence of the remarkable fact that measurement channels -- while being
entanglement-breaking -- can show unlimited gain of entanglement assistance
in the classical capacity \cite{h5}. At this point we would like to add that
recently a quantum communication scheme was proposed that utilizes
pre-shared CV entanglement, and in principle can demonstrate theoretically
predicted capacity enhancement for noisy quantum attenuator channel \cite%
{guha}. It would be worthwhile to investigate designs which can achieve a
similar goal for entanglement-assisted quantum CV measurements.

The plan of the paper is as follows: in sec. \ref{s2} we recall the notion
of measurement channel and its entropy reduction. In sec. \ref{s3} we
briefly describe the protocol of entanglement-assisted measurement and
summarize in theorem \ref{t1} relevant results from our papers \cite{H1},
\cite{H2} concerning entanglement-assisted capacity. A detailed proof of the
main result concerning the extremal property of the entropy reduction of
Gaussian measurement channel is given in sec. \ref{s4} in the case of the
global gauge symmetry; then the gain of entanglement assistance is
demonstrated on the example of one mode in sec. \ref{s5}. We have chosen to
consider first the phase insensitive case because it is of special
importance in applications while admitting relatively direct treatment and
transparent description. Finally, the extension of the basic extremal
property to the case of general Gaussian observables is outlined in sec. \ref%
{s6}.

\section{Entropy reduction of a measurement channel}

\label{s2}

Let $\mathcal{H}$ be a separable Hilbert space of a quantum system, $%
\mathfrak{S}(\mathcal{H})$ the set of all density operators (quantum
states), and let $(\Omega ,\mathcal{F},\mu ),$ 
be a standard measurable space, where $\mu $ is a $\sigma $-finite measure
on the $\sigma $-algebra $\mathcal{F}$.

\emph{Quantum observable} with values in $\Omega $ is a probability
operator-valued measure (POVM) $M=\{M(A),A\in \mathcal{F}\}$ on $(\Omega ,
\mathcal{F}).$ \emph{The probability distribution} of observable $M$ in the
state $\rho \in \mathfrak{S}(\mathcal{H})$ is given by the formula
\begin{equation*}
P_{\rho }(A)=\mathrm{Tr}\rho M(A),\quad A\in \mathcal{F}.
\end{equation*}%
\emph{Measurement channel} $\mathcal{M}$ is an affine map $\rho \rightarrow
P_{\rho }(d\omega )$ of the convex set of quantum states $\mathfrak{S}(%
\mathcal{H})$ into the set of probability distributions on $\Omega .$

We will deal with the special class of observables which have \emph{bounded
operator valued density} i.e.%
\begin{equation}
M(A)=\int_{A}m({\omega })\,\mu (d\omega ),\quad A\in \mathcal{F},
\label{den}
\end{equation}%
where $m({\omega })=V({\omega })^{\ast }V({\omega })$ and $V({\omega })$ is
weakly measurable function with values in the algebra of bounded operators
in $\mathcal{H}$ such that%
\begin{equation*}
\int_{\Omega }V({\omega })^{\ast }V({\omega })\mu (d\omega )=I,
\end{equation*}%
and the integral weakly converges. With any measurable factorization of $m({%
\omega })$ one can associate an \emph{efficient instrument} (see \cite{Oz},
\cite{Sh-ERQM}) defined by the probability distribution $P_{\rho }(d\omega )$
with the density
\begin{equation}
p_{\rho }(\omega )=\mathrm{Tr}\rho V({\omega })^{\ast }V({\omega })
\label{opd}
\end{equation}%
with respect to the measure $\mu ,$ and by the family of posterior states
\begin{equation*}
\hat{\rho}(\omega )=%
\begin{cases}
(p_{\rho }(\omega ))^{-1}V(\omega )\rho V(\omega )^{\ast }, & \text{ if $%
p_{\rho }(\omega )\neq 0$;} \\
\hat{\rho}_{0}, & \text{otherwise}%
\end{cases}%
\end{equation*}%
for any state $\rho \in \mathfrak{S}(\mathcal{H})$ ($\hat{\rho}_{0}$ is a
fixed state).

Folowing the \cite{Oz}, \cite{Sh-ERQM}, one defines the \emph{entropy
reduction} of the efficient instrument by the formula
\begin{equation}
ER(\rho ,\mathcal{M})=\mathrm{H}(\rho )-\int_{\Omega }p_{\rho }(\omega )\,%
\mathrm{H}(\hat{\rho}(\omega ))\mu (d\omega ),  \label{er-1}
\end{equation}%
where $\mathrm{H}(\rho )=-\mathrm{Tr}\rho \log \rho ,$ provided $\mathrm{H}%
(\rho )<\infty .$ In \cite{Oz} it was shown that the entropy reduction of an
efficient instrument is nonnegative. In \cite{Sh-ERQM} the entropy reduction
was related to quantum mutual information of the instrument and hence it is
a concave, subadditive, lower semicontinuous function of $\rho .$

Let us make an important observation: the entropies of posterior states
depend only on $m({\omega })$ and not on the way of measurable factorization
$m({\omega })=V({\omega })^{\ast }V({\omega })$ because trace-class operator
$V(\omega )\rho V(\omega )^{\ast }=\left( V(\omega )\sqrt{\rho }\right)
\left( V(\omega )\sqrt{\rho }\right) ^{\ast }$ has the same eigenvalues as $%
\left( V(\omega )\sqrt{\rho }\right) ^{\ast }\left( V(\omega )\sqrt{\rho }%
\right) =\sqrt{\rho }\,m(\omega )\sqrt{\rho }.$ Thus the entropy reduction (%
\ref{er-1}) is uniquely defined for any observable $M$ of the form (\ref{den}%
), which justifies the notation $ER(\rho ,\mathcal{M}).$

Let $\mathfrak{B}(\mathcal{H})$ be the algebra of all bounded operators and $%
\mathfrak{T}(\mathcal{H})$ the Banach space of trace-class operators in $%
\mathcal{H}$. A hybrid classical-quantum (cq) system \cite{BL-2} is
described by the von Neumann algebra $\mathcal{L}=\mathcal{L}^{\infty
}(\Omega ,\mathcal{F},\mu ,\mathfrak{B}(\mathcal{H}))$, consisting of weakly
measurable, essentially bounded functions $X(\omega ),\omega \in \Omega $
with values in $\mathfrak{B}(\mathcal{H})$. The elements of the preadjoint
space $\mathcal{L}_{\ast }=\mathcal{L}_{1}(\Omega ,\mathcal{F},\mu ,%
\mathfrak{T}(\mathcal{H}))$ are measurable functions $\varrho =\{\rho
(\omega )\}$ with values in $\mathfrak{T}(\mathcal{H}),$ integrable with
respect to the measure $\mu $. An element $\varrho =\{\rho (\omega )\}\in
\mathcal{L}_{\ast },$ such that
\begin{equation*}
\rho (\omega )\geq 0~(mod~\mu ),~\int_{\Omega }\mathrm{Tr}\rho (\omega )\mu
(d\omega )=1,
\end{equation*}%
is called \emph{cq-state}. The partial $c$-state is the probability measure
on $(\Omega ,\mathcal{F}),$ determined by the density $p(\omega )=\mathrm{Tr}%
\rho (\omega )$ with respect to the measure $\mu (d\omega )$; the partial $q$%
-state is the density operator $\rho =\int_{\Omega }\rho (\omega )\mu
(d\omega ).$

Let $\mathcal{M}$ be a measurement channel introduced above, then the
relation $\varrho =\{V(\omega )\rho V(\omega )^{\ast }\}$ defines a
cq-state. The map $\rho \rightarrow \mathcal{E}[\rho ]=\left\{ V(\omega
)\rho V(\omega )^{\ast }\right\} $ is a channel with quantum input and
cq-output.

\section{Entanglement-assisted capacity of a measurement channel}

\label{s3}

In the ordinary (unassisted) measurement scenario there are two parties --
quantum system $A$ (the measured system), classical system $\Omega $ (the
meter), and the measurement channel $\mathcal{M}:A\rightarrow \Omega .$ In
the case of infinite-dimensional $\mathcal{H}$ one usually introduces energy
constraint $\mathrm{Tr}\rho H\leq E$ onto the input states $\rho $ of the
channel. Here $H$ is positive selfadjoint (in general unbounded)
\textquotedblleft energy operator\textquotedblright\ (Hamiltonian) on the
space $\mathcal{H}$ of the system $A,$ $E$ is a positive constant, and the
trace is understood e.g. as in sec. 11.1 of \cite{QSCI}. 
A natural (but not the unique) measure of information-processing performance
of the measurement is the energy-constrained classical capacity $C(\mathcal{M%
},H,E)$ of the channel $\mathcal{M}$, see \cite{H2}.

The protocol of entanglement-assisted classical communication via
finite-dimensional quantum channel was introduced in \cite{bsst1}, \cite%
{BSST}. A modification of this protocol for quantum measurement channels,
which requires the notion of hybrid $cq$ system, was studied in \cite{H1},
\cite{H2}.

We give here a brief description of the \emph{entanglement-assisted
measurement} protocol including the resulting capacity formula (\ref{cmf})
which is sufficient for our purposes. In this scenario the meter is a
classical-quantum system $\Omega B$ where $B$ is its quantum part. The
composite quantum system $AB$ is initially in a pure entangled state $\rho
_{AB}$. The party $A$ performs encoding $x\rightarrow {\mathcal{E}}_{x}$ of
the classical signal$~x$, 
where ${\mathcal{E}}_{x}$ are operations on the measured system. Thereafter
the measurement channel $\mathcal{M}:A\rightarrow \Omega $ is applied so
that the meter $\Omega B$ is transformed into one the $cq$-states $(\mathcal{%
M}\circ {\mathcal{E}}_{x}\otimes \mathrm{Id}_{B})[\rho _{AB}].$ The goal is
to extract the maximum information about $x$ basing on measurements in the
hybrid system $\Omega B$. With the block coding, this procedure should be
applied to the channel $\mathcal{M}^{\otimes n}$ whose input states satisfy
the corresponding energy constraint. 
The asymptotic (as $n\rightarrow \infty $) capacity of this protocol is
called the \emph{energy-constrained classical entanglement-assisted capacity}
$C_{ea}(\mathcal{M},H,E)$ of the measurement channel $\mathcal{M}$. We refer
to \cite{H1}, \cite{H2} for explanation of the relevant details.

In the following theorem we summarize the relevant results from \cite{H2}
and \cite{QSCI} giving a convenient expression for $C_{ea}(\mathcal{M},H,E)$
in terms of the entropy reduction.

\begin{theorem}
\label{t1} \emph{Let $\mathcal{M}$ be a measurement channel with observable
of the form (\ref{den}) such that}%
\begin{equation}
\sup_{\rho :~\mathrm{Tr}\rho H\leq E}\mathrm{H}_{c}(\mathcal{M}(\rho
))<\infty ,  \label{suph}
\end{equation}%
\emph{where }$\mathrm{H}_{c}$\emph{$(\mathcal{M}(\rho ))$ is the classical
differential entropy of the output probability density (\ref{opd}) of the
channel.}

\emph{Assume that the energy operator $H$ satisfies the condition
\begin{equation}
\mathrm{Tr}\exp (-\beta H)<\infty \mbox{ for all }\beta >0.  \label{F_cond1}
\end{equation}%
}

\emph{Then the energy-constrained entanglement-assisted capacity is finite
and is given by the formula
\begin{equation}
C_{ea}(\mathcal{M},H,E)=\max_{\rho :~\mathrm{Tr}\rho H\leq E}ER(\rho ,%
\mathcal{M}).  \label{cmf}
\end{equation}%
Moreover, if the channel $\mathcal{M}$ is such that
\begin{equation}
\sup_{\rho }ER(\rho ,\mathcal{M})=+\infty ,  \label{eqa}
\end{equation}%
then the maximum in (\ref{cmf}) is achieved on the density operator $\rho $,
such that $\mathrm{Tr}\rho H = E$.}
\end{theorem}

\emph{Proof (sketch).} By theorem 3 of \cite{H1}, the conditions (\ref{suph}%
), (\ref{F_cond1}) imply%
\begin{equation*}
C_{ea}(\mathcal{M},H,E)=\sup_{\rho :~\mathrm{Tr}\rho H\leq E}ER(\rho ,%
\mathcal{M}).
\end{equation*}%
Next notice that the quantum entropy $\mathrm{H}(\rho )$ is bounded and
continuous on the set $\left\{ \rho :~\mathrm{Tr}\rho H\leq E\right\} $ by
lemma 11.8 of \cite{QSCI}, provided the operator $H$ satisfies the condition
(\ref{F_cond1}). This implies that $ER(\rho ,\mathcal{M})$ is well-defined
and also continuous by theorem 2 of \cite{Sh-ERQM}. The set $\left\{ \rho :~%
\mathrm{Tr}\rho H\leq E\right\} $ is compact by lemma 11.5 of \cite{QSCI},
hence the supremum of the entropy reduction is achieved on this set, and the
formula (\ref{cmf}) holds.

By using the concavity of the entropy reduction, the second statement can be
proved similarly to the corresponding statement of proposition 11.26 of \cite%
{QSCI} . $\square $

In the recent paper \cite{acc-noJ} it was shown that the conditions (\ref%
{suph}) and (\ref{F_cond1}) are fulfilled in the case of Gaussian
measurement channel with the constraint given by an oscillator-system
Hamiltonian. Thus the relation (\ref{cmf}) holds in this case to which we
pass in the next section.

\section{Gauge-covariant Gaussian measurements}

\label{s4}

In what follows $\mathcal{H}$ will be the space of a strongly continuous
irreducible representation of bosonic canonical commutation relations (CCR)
(see e.g. \cite{asp}, \cite{QSCI} for a detailed account) describing
quantization of a linear classical system with $s$ degrees of freedom such
as finite number of physically relevant electromagnetic modes in a
receiver's cavity. Let $a_{j},a_{j}^{\dagger };j=1,\dots ,s$ be the
annihilation/creation operators of the modes, let $\ z\in \mathbb{C}^{s}$ be
a column vector with complex coordinates $z_{j},j=1,\dots ,s,$ and $z^{\ast
} $ denote Hermitian conjugate row vector. Then the CCR are conveniently
written in terms of the unitary displacement operators $D(z)=\exp
\sum\limits_{j=1}^{s}\left( a_{j}^{\dagger }\,z_{j}-\bar{z}%
_{j}\,a_{j}\right) ,$ namely
\begin{equation}
D(z)D(w)=\exp \left( -i\,\mathrm{Im\,}z^{\ast }w\right) D(z+w),\quad z,w\in
\mathbb{C}^{s}.  \label{ccr}
\end{equation}

The (global) gauge group acts as $z\rightarrow e^{i\varphi }z,$ ($\varphi $
is real phase) in the space $\mathbb{C}^{s}$, and via the unitary group $%
\varphi \rightarrow U_{\varphi }=\exp \left( -i\varphi \,\mathcal{N}\right) $
in $\mathcal{H}$ (here $\mathcal{N\,}$=$\sum\limits_{j=1}^{s}a_{j}^{\dagger
}\,a_{j}$ is the total number operator), so that%
\begin{equation*}
U_{\varphi }^{\ast }D(z)U_{\varphi }=D(e^{i\varphi }z).
\end{equation*}%
An operator $A$ is gauge-invariant if $U_{\varphi }AU_{\varphi }^{\ast }=A$
for all $\varphi .$ A gauge-invariant Gaussian state has the quantum
characteristic function \cite{asp}
\begin{equation}
\mathop{\rm Tr}\nolimits\rho _{\Lambda }D(w)=\exp \left[ -w^{\ast }\left(
\Lambda +\frac{I_{s}}{2}\right) w\right] ,  \label{one-mode-cf}
\end{equation}%
where $\Lambda =\mathrm{Tr}\,a\rho _{\Lambda }a^{\dagger }$ is the complex
correlation matrix, satisfying $\Lambda \geq 0.$ (We denote by $I_{s}$ the
unit $s\times s$-matrix, as distinct from the unit operator $I$ in a Hilbert
space). The case $\Lambda =0$ in (\ref{one-mode-cf}) corresponds to the
vacuum state $\rho _{0}=|0\rangle \langle 0|.$ The coherent state vectors
are $|z\rangle =D(z)|0\rangle .$


The displaced state $\rho _{\Lambda ,z}= D(z)\rho _{\Lambda }D(z)^*$ has the
quantum characteristic function
\begin{equation}
\mathop{\rm Tr}\nolimits\rho _{\Lambda ,z}D(w)=\exp \left[ 2i\mathrm{Im\,}%
z^{\ast }w-w^{\ast }\left( \Lambda +\frac{I_{s}}{2}\right) w\right] .
\label{lamz}
\end{equation}
We will use the P-representation in the case of nondegenerate $\Lambda :$
\begin{equation}
\rho _{\Lambda ,z}\equiv D(z)\rho _{\Lambda }D(z)^*=\int |w\rangle \langle
w|\exp \left( -\left( w-z\right) ^{\ast }\Lambda ^{-1}\left( w-z\right)
\right) \frac{d^{2s}w}{\pi ^{s}\det \Lambda }.  \label{shifted2}
\end{equation}
The formula (\ref{lamz}) remains valid for arbitrary correlation matrix $%
\Lambda \geq 0,$ while (\ref{shifted2}) needs modification by introducing
Gaussian measure on $\mathbb{C}^{s}$ with zero mean and complex correlation
matrix $\Lambda .$ For the sake of clarity, we will deal with the case of
nondegenerate $\Lambda ,$ while the resulting formulas remain valid for
arbitrary $\Lambda \geq 0.$

In this section we will consider the gauge-covariant Gaussian observable
(POVM) with values in $\Omega =\mathbb{C}^{s}$ defined by
\begin{equation}
M(d^{2s}z)=D(z)\rho _{N}D(z)^{\ast }\frac{d^{2s}z}{\pi ^{s}},
\label{gauss_POVM}
\end{equation}%
where $N\geq 0$ is the correlation matrix of the measurement noise. The case
$N=0 $ corresponds to the multimode heterodyne measurement (see \cite{acc}
for more detail). Put $\mu (dz)=\frac{d^{2s}z}{\pi ^{s}},$ then observable\ (%
\ref{gauss_POVM}) has the form (\ref{den}) with $m(z)=D(z)\rho
_{N}D(z)^{\ast }$\ taking values in the space of trace-class operators. For
any input state $\rho $ the output probability density (\ref{opd}) of the
corresponding measurement channel $\mathcal{M}$ is
\begin{equation}  \label{density}
p_{\rho }(z)=\mathrm{Tr}\rho D(z)\rho _{N}D(z)^{\ast }.
\end{equation}
Choosing $V(z)=\sqrt{\rho _{N}}D(z)^{\ast },$ we get the posterior states
\begin{equation}
\hat{\rho}(z)=p_{\rho }(z)^{-1}V(z)\,\rho \,V(z)^{\ast }=p_{\rho }(z)^{-1}%
\sqrt{\rho _{N}}D(z)^{\ast }\rho D(z)\sqrt{\rho _{N}},  \label{gpost}
\end{equation}%
and the entropy reduction is given by (\ref{er-1}).

Assume that
\begin{equation}
H=\sum_{j,k=1}^{s}\epsilon _{jk}\,a_{j}^{\dagger }a_{k}  \label{haml}
\end{equation}%
is a quadratic gauge-invariant oscillator-type Hamiltonian, where $\epsilon =%
\left[ \epsilon _{jk}\right] $ is positive definite Hermitian matrix. In
\cite{acc-noJ} we have shown that in the case of observable (\ref{gauss_POVM}%
) and the Hamiltonian (\ref{haml}) the conditions (\ref{suph}) and (\ref%
{F_cond1}) are fulfilled, making the formula (\ref{cmf}) for $C_{ea}(%
\mathcal{M};H,E)$ applicable.

Throughout this paper we use the fact that for any state $\rho $ with finite
second moments $\mathrm{H}(\rho )$ is finite (it is upperbounded by the
entropy of the Gaussian state with the same second moments), hence the
entropy reduction is well-defined.

\begin{proposition}
\label{p1} \emph{Let $\mathcal{M}$ be the measurement channel corresponding
to the observable (\ref{gauss_POVM}). Then for any state $\rho $ with finite
second moments there is a gauge-invariant state $\rho _{gi}$ such that}
\begin{equation}
ER(\rho ,\mathcal{M})\leq ER(\rho _{gi},\mathcal{M});\quad \mathrm{Tr}\rho H=%
\mathrm{Tr}\rho _{gi}H.  \label{ergi}
\end{equation}
\end{proposition}

The \emph{proof} is similar to that of Corollary 12.39 in \cite{QSCI}.
Define the gauge-invariant state
\begin{equation*}
\rho _{gi}=\int\limits_{0}^{2\pi }U_{\varphi }^{\ast }\rho U_{\varphi }\frac{%
d\varphi }{2\pi },
\end{equation*}%
then the second relation in (\ref{ergi}) follows from the fact that $%
U_{\varphi }HU_{\varphi }^{\ast }=H,$ and the first one -- from Jensen's
inequality relying upon nonnegativity, concavity and lower semicontinuity of
$ER(\rho ,\mathcal{M})$ \cite{shir}. $\square $

By $\mathfrak{S}(\Lambda )$ we denote the set of all states which have
finite second moments with the \emph{complex correlation matrix}
\begin{equation*}
\left[ \mathrm{Tr}\,a_{j}\rho a_{k}^{\dagger }\right] _{j,k=1,\dots
,s}=\Lambda .
\end{equation*}%
If $\rho \in \mathfrak{S}(\Lambda ),$ then $\rho _{gi}\in \mathfrak{S}%
(\Lambda )$ and it has zero first moments, and second moments such as $%
\mathrm{Tr}\,a_{j}\rho_{gi} a_{k},\,\mathrm{Tr\,}a_{j}^{\dagger }\rho_{gi}
a_{k}^{\dagger }$ vanishing. This follows from the identities $U_{\varphi
}^{\ast }\,a_{j}U_{\varphi }=a_{j}e^{-i\varphi },\quad U_{\varphi }^{\ast
}\,a_{j}^{\dagger }U_{\varphi }=a_{j}^{\dagger }e^{i\varphi }.$ Moreover the
\emph{normal} second moments such as $\mathrm{Tr}\,a_{j}\rho a_{k}^{\dagger
} $ coincide for $\rho $ and $\rho _{gi}$. There is a unique gauge-invariant
Gaussian state $\rho _{\Lambda }$ in $\mathfrak{S}(\Lambda )$ and the \emph{%
normal} second moments coincide for $\rho \in \mathfrak{S}(\Lambda )$ and $%
\rho _{\Lambda}. $

We will study the following quantity
\begin{equation}
ER(\mathcal{M};\Lambda )=\sup_{\rho \in \mathfrak{S}(\Lambda )}ER(\rho ,%
\mathcal{M}).  \label{cchi}
\end{equation}%
This quantity, which is interesting on its own, is of the main importance in
computing the energy-constrained classical entanglement-assisted capacity of
the measurement channel. Indeed, assume that the Hamiltonian is given by (%
\ref{haml}), so that the mean energy of the input state $\rho $ is equal to%
\begin{equation*}
\mathrm{Tr}\rho H=\sum_{j,k=1}^{s}\epsilon _{jk}\Lambda _{kj}=\mathrm{Sp\,}%
\epsilon \Lambda ,
\end{equation*}%
where $\mathrm{Sp}$ denotes trace of $s\times s-$matrices as distinct from
the trace of operators. Then the energy constraint has the form $\mathrm{Sp\,%
}\epsilon \Lambda \leq E,$ and according to (\ref{cmf}) the
energy-constrained entanglement-assisted classical capacity of the channel $%
\mathcal{M}$ is
\begin{equation}
C_{ea}(\mathcal{M};H,E)=\max_{\Lambda :\mathrm{Sp\,}\epsilon \Lambda \leq
E}ER(\mathcal{M};\Lambda ).  \label{ceag}
\end{equation}%
Given an explicit expression for $ER(\mathcal{M};\Lambda )$ such as (\ref%
{erfin}) below, computation of the last supremum is a separate optimization
problem. Moreover, from (\ref{erfin}) it follows that $\sup_{\Lambda
}ER(\rho _{\Lambda },\mathcal{M})=+\infty ,$ which implies the condition (%
\ref{eqa}) in theorem \ref{t1}. Hence the maximum in (\ref{ceag}) is
attained on a $\Lambda $ satisfying $\mathrm{Sp\,}\epsilon \Lambda =E.$

\begin{theorem}
\label{t2} \emph{Let $\mathcal{M}$ be the measurement channel corresponding
to the observable (\ref{gauss_POVM}). Then the maximum of entropy reduction $%
ER(\rho ,M)$ on $\mathfrak{S}(\Lambda )$ is attained on the gauge-invariant
Gaussian state $\rho _{\Lambda }.$ Moreover,}%
\begin{equation}
ER(\mathcal{M};\Lambda )=\mathrm{Sp}\,\,g(\Lambda )-\mathrm{Sp}\,g(\tilde{N}%
),  \label{erfin}
\end{equation}%
\emph{where $g(x)=(x+1)\log (x+1)-x\log x$, and}
\begin{equation}
\tilde{N}=\sqrt{N(N+I_{s})^{-1}}\Lambda (\Lambda +N+I_{s})^{-1}\sqrt{N\left(
N+I_{s}\right) }.  \label{tildeN}
\end{equation}
\end{theorem}

\emph{Proof.} Due to proposition \ref{p1}, in consideration of the maximum
of $ER(\rho ,M)$ we can restrict to gauge-invariant states $\rho \in
\mathfrak{S}(\Lambda )$. Denote $V(z)=\sqrt{\rho _{N}}D(z)^{\ast }$ and
consider the $cq$-states
\begin{eqnarray*}
\varrho &=&\{\rho (z)\},\,\quad \rho (z)=V(z)\,\rho \,V(z)^{\ast }, \\
\varrho _{\Lambda } &=&\{\rho _{\Lambda }(z)\},\,\quad \rho _{\Lambda
}(z)=V(z)\,\rho _{\Lambda }\,V(z)^{\ast },
\end{eqnarray*}%
then the $c$-states $P$ and $P_{\Lambda }$ are defined by densities $p(z)=%
\mathrm{Tr}\rho (z)$ and $p_{\Lambda }(z)=\mathrm{Tr}\rho _{\Lambda }(z)$.
We have
\begin{eqnarray}
&&ER(\rho _{\Lambda },\mathcal{M})-ER(\rho ,\mathcal{M})  \label{difer} \\
&=&\mathrm{H}(\rho \Vert \rho _{\Lambda })-\mathrm{H}_{cq}(\varrho \Vert
\varrho _{\Lambda })\,\,+\mathrm{H}_{c}(P\Vert P_{\Lambda })  \notag \\
&+&\mathrm{Tr}(\rho -\rho _{\Lambda })\log \rho _{\Lambda }+\int \mathrm{Tr}%
(\rho _{\Lambda }(z)-\rho (z))\log \,\hat{\rho}_{\Lambda }(z)\frac{d^{2s}z}{%
\pi ^{s}},  \notag
\end{eqnarray}%
where $\hat{\rho}_{\Lambda }(z)=\rho _{\Lambda }(z)/p_{\Lambda }(z)$ are the
posterior states corresponding to the input state $\rho _{\Lambda },$
\begin{equation*}
\mathrm{H}_{c}(P\Vert P_{\Lambda })=\int p(z)\log \left( \frac{p(z)}{%
p_{\Lambda }(z)}\right) \frac{d^{2s}z}{\pi ^{s}}
\end{equation*}%
is the classical relative entropy of $P$, $P_{\Lambda }$ and
\begin{equation*}
\mathrm{H}_{cq}(\varrho \Vert \varrho _{\Lambda })=\int \mathrm{Tr\,}\rho
(z)\,\left( \log \rho (z)-\log \rho _{\Lambda }(z)\right) \frac{d^{2s}z}{\pi
^{s}}=\mathrm{H}(\mathcal{E}\,[\rho ]\Vert \mathcal{E}\,[\rho _{\Lambda }]),
\end{equation*}%
is the relative entropy of $cq$-states (see Eq. (3) in \cite{BL-2}). Here we
use the channel $\mathcal{E}\,[\rho ]=\,\left\{ V(z)\,\rho \,V(z)^{\ast
}\right\} $ with quantum input and hybrid $cq$ output.

Monotonicity of the relative entropy for $cq$-states (\cite{BL-2}, theorem
1) then implies
\begin{equation*}
\mathrm{H}_{cq}(\varrho \Vert \varrho _{\Lambda })\leq \mathrm{H}(\rho \Vert
\rho _{\Lambda }),
\end{equation*}%
hence we have for the first three terms in (\ref{difer})
\begin{equation}
\mathrm{H}(\rho \Vert \rho_{\Lambda })-\mathrm{H}_{cq}(\varrho \Vert \varrho
_{\Lambda })\,+\mathrm{H}_{c}(P\Vert P_{\Lambda })\geq 0.  \label{first}
\end{equation}

As we have assumed, $\Lambda $ is non-degenerate hence $\log \rho _{\Lambda
} $ exists and is a linear combination of the operators $I,a_{j}^{\dagger
}a_{k}.$ This follows from the exponential form of the density operator
(theorem 12.23 in \cite{QSCI}). Since $\rho ,\rho _{\Lambda }\in \mathfrak{S}%
(\Lambda ),$ then the normal second moments of the states $\rho $ and $\rho
_{\Lambda }$ coincide and hence
\begin{equation}
\mathrm{Tr}(\rho -\rho _{\Lambda })\log \rho _{\Lambda }=0.  \label{second}
\end{equation}%
It remains to show that also
\begin{equation}
\int \mathrm{Tr}(\rho _{\Lambda }(z)-\rho (z))\log \,\hat{\rho}_{\Lambda
}(z)\,\frac{d^{2s}z}{\pi ^{s}}=0.  \label{res}
\end{equation}

\begin{lemma}
\label{l1} \emph{Let $\rho =\rho _{\Lambda }$ then the posterior state (\ref%
{gpost}) is the Gaussian state}
\begin{equation}
\hat{\rho}_{\Lambda }(z)= D(Kz)^{\ast }\rho _{\tilde{N}}D(Kz),
\label{poster}
\end{equation}%
where%
\begin{eqnarray*}
K &=&\sqrt{N\left( N+I_{s}\right) }(\Lambda +N+I_{s})^{-1}, \\
\tilde{N} &=&\sqrt{N\left( N+I_{s}\right) ^{-1}}\Lambda (\Lambda
+N+I_{s})^{-1}\sqrt{N\left( N+I_{s}\right) }.
\end{eqnarray*}
\end{lemma}

\emph{Proof}. By using the quantum Parceval relation%
\begin{equation}
\mathop{\rm Tr}\nolimits\rho \sigma ^{\ast }=\int \mathop{\rm Tr}
\nolimits\rho D(w)\,\overline{\mathop{\rm Tr}\nolimits\sigma D(w)}\frac{
d^{2s}w}{\pi ^{s}},  \label{parc}
\end{equation}%
the relation (\ref{density}) for $\rho=\rho_\Lambda$ and the characteristic
functions of the Gaussian states, we can show as in \cite{acc} that
\begin{equation*}
p_{\Lambda }(z)=\frac{1}{\det \left( \Lambda +N+I_{s}\right) }\exp \left(
-z^{\ast }(\Lambda +N+I_{s})^{-1}z\right) .
\end{equation*}

It is known (see \cite{tmf}) that the square root of a Gaussian density
operator is proportional to another Gaussian density operator. By using the
Fock basis in $\mathcal{H}$ associated with the eigenvectors of the matrix $%
N $ (see e.g. Appendix in \cite{acc}), we obtain
\begin{equation}
\sqrt{\rho _{N}}=c\rho _{L},~\quad L=N+\sqrt{N\left( N+I_{s}\right) },
\label{square}
\end{equation}%
\begin{equation*}
~c^{2}=\det (2L+I_{s})=\det \left( \sqrt{N}+\sqrt{N+I_{s}}\right) ^{2}.
\end{equation*}%
A calculation in this Fock basis shows also that, similarly to Eq. (33) of
\cite{acc},
\begin{equation}
\sqrt{\rho _{N}}\ket w=\frac{1}{\det \sqrt{N+I_{s}}}\exp \left( -\frac{%
w^{\ast }(N+I_{s})^{-1}w}{2}\right) \,\left\vert {\sqrt{N(N+I_{s})^{-1}}w}%
\right\rangle ,  \label{omega1}
\end{equation}%
whence, by using the P-representation (\ref{shifted2}) for $\rho _{\Lambda
,-z}=D(z)^{\ast }\rho _{\Lambda }D(z)$
\begin{eqnarray}
\hat{\rho}_{\Lambda }(z) &=&p_{\Lambda }(z)^{-1}\sqrt{\rho _{N}}\rho
_{\Lambda ,-z}\sqrt{\rho _{N}}  \notag \\
&=&\frac{p_{\Lambda }(z)^{-1}}{\det \Lambda {(N+I_{s})}}\int \left\vert {%
\sqrt{N(N+I_{s})^{-1}}w}\right\rangle \left\langle {\sqrt{N(N+I_{s})^{-1}}w}%
\right\vert  \notag \\
&\times &\exp \left( -w^{\ast }(N+I_{s})^{-1}w-\left( w+z\right) ^{\ast
}\Lambda ^{-1}\left( w+z\right) \right) \frac{d^{2s}w}{\pi ^{s}}  \notag \\
&=&\frac{1}{\det \tilde{N}}\int \ketbra uu\exp \left( -\left( u+Kz\right)
^{\ast }\tilde{N}^{-1}\left( u+Kz\right) \right) \frac{d^{2s}u}{\pi ^{s}}
\notag \\
&=&D(Kz)^{\ast }\rho _{\tilde{N}}D(Kz),  \label{M-apost}
\end{eqnarray}%
where we made the change of variable $u=\sqrt{N(N+I_{s})^{-1}}w$. $\square $

Substituting the posterior state (\ref{poster}) into the right-hand side of (%
\ref{res}), we obtain
\begin{equation}  \label{res3}
\int \mathrm{Tr}(\rho _{\Lambda }(z)-\rho (z))\log \hat{\rho}_{\Lambda }(z)%
\frac{d^{2s}z}{\pi ^{s}} = \mathrm{Tr}\,\left[ \Phi _{M}\,(\rho _{\Lambda
})-\Phi _{M}(\rho )\right] \log \rho _{\tilde{N}},
\end{equation}
where we have introduced the channel
\begin{equation}
\Phi _{M}(\sigma )=\int D(Kz)\sqrt{\rho _{N}}\,D(z)^{\ast }\,\sigma \,D(z)%
\sqrt{\rho _{N}}\,D(Kz)^{\ast }\frac{d^{2s}z}{\pi ^{s}}.  \label{M-add}
\end{equation}

\begin{lemma}
\label{l2} \emph{$\Phi _{M}$ is a gauge-covariant Gaussian channel.}
\end{lemma}

We will give the proof in a moment, but first let us explain how this lemma
implies the required identity (\ref{res}). The state $\rho \in \mathfrak{S}%
(\Lambda )$ and $\rho _{\Lambda }$ have the same normal second moments,
which are transformed similarly under the action of a gauge-covariant\emph{\
}Gaussian channel. Hence $\Phi _{M}(\rho )$ and $\Phi _{M}\,(\rho _{\Lambda
})$ also have the same normal second moments. Without loss of generality, we
can assume that $N$ hence $\tilde{N}$ is nondegenerate. Then $\log \rho _{%
\tilde{N}}$ \ exists and is a linear combination of the operators $%
I,a_{j}^{\dagger }a_{k},$ hence
\begin{equation*}
\mathrm{Tr}\,\left[ \Phi _{M}\,(\rho _{\Lambda })-\mathrm{Tr}\,\Phi
_{M}(\rho )\right] \log \rho _{\tilde{N}}=\mathrm{Tr}\,\,(\rho _{\Lambda
}-\rho )\,\Phi _{M}^{\ast }\left( \log \rho _{\tilde{N}}\right) =0.
\end{equation*}
Taking into account (\ref{res3}) this implies (\ref{res}). Together with (%
\ref{first}) and (\ref{second}) this implies that the difference (\ref{difer}%
) is nonnegative, i.e. the first statement of the theorem. The formula (\ref%
{erfin}) follows from
\begin{equation*}
ER(\rho _{\Lambda },\mathcal{M})=\mathrm{H}(\rho _{\Lambda })-\int
p_{\Lambda }(z)\mathrm{H}(\hat{\rho}_{\Lambda }(z))\frac{d^{2s}z}{\pi ^{s}},
\end{equation*}%
and the fact that $\mathrm{H}(\hat{\rho}_{\Lambda }(z))=\mathrm{H}(\rho _{%
\tilde{N}})$ because of the unitary equivalence (\ref{poster}).

It remains to prove the lemma \ref{l2}. We will do this by checking the
definition of the dual Gaussian channel (see \cite{QSCI}, sec. 12.4.2). We
have
\begin{equation*}
\Phi _{M}^{\ast }(D(w))=\int \exp (-2i\,\mathrm{Im}(w^{\ast }Kz))D(z)\sigma
D(z)^{\ast }\frac{d^{2s}z}{\pi ^{s}}\equiv X,
\end{equation*}%
where%
\begin{equation*}
~\sigma =\sqrt{\rho _{N}}\,D(w)\sqrt{\rho _{N}}.
\end{equation*}%
One has $X=\left[ \mathrm{Tr}\sigma D(K^{\ast }w)^{\ast }\right] D(K^{\ast
}w),$ indeed by using (\ref{ccr})%
\begin{equation*}
X\,D(K^{\ast }w)^{\ast }=\int D(z)\sigma D(K^{\ast }w)^{\ast }D(z)^{\ast }%
\frac{d^{2s}z}{\pi ^{s}}=\mathrm{Tr}\left( \sigma D(K^{\ast }w)^{\ast
}\right) \,I,
\end{equation*}%
where the second equality follows from the orthogonality relations for the
irreducible representation $z\rightarrow D(z)$ (see e.g. sec. I.3.5 of \cite%
{asp}). Thus
\begin{equation}
\Phi _{M}^{\ast }(D(w))=\varphi (w)D(K^{\ast }w),~  \label{gd}
\end{equation}%
where $\varphi (w)=\mathrm{Tr}\sqrt{\rho _{N}}\,D(w)\sqrt{\rho _{N}}%
D(K^{\ast }w)^{\ast }.$

By using (\ref{square}) and the quantum Parceval relation (\ref{parc}) we
have
\begin{equation}
\varphi (w)=c^{2}\int \mathrm{Tr}\rho _{L}D(w)D(z)\,\overline{\mathrm{Tr}%
D(K^{\ast }w)\rho _{L}D(z)}\,\frac{d^{2s}z}{\pi ^{s}}.  \label{varphi1}
\end{equation}%
Denote $R=2L+I_{s}.$ Then (\ref{ccr}) and the formula for characteristic
function of Gaussian state imply that (\ref{varphi1}) is equal to
\begin{eqnarray}
&&c^{2}\int \exp (i\,\mathrm{Im}{z^{\ast }}(I_{s}+K^{\ast })w)\,\mathrm{Tr}%
\rho _{L}D(w+z)\overline{\mathrm{Tr}\rho _{L}D(K^{\ast }w+z})\frac{d^{2s}z}{%
\pi ^{s}}  \notag \\
&=&c^{2}\int \exp \left( i\,\mathrm{Im}{z^{\ast }}(I_{s}+K^{\ast })w-z^{\ast
}R\,z-\mathrm{Re}\,z^{\ast }R\,w\right)  \notag \\
&\times &\exp \left( -\mathrm{Re}\,z^{\ast }R\,K^{\ast }w-\frac{1}{2}w^{\ast
}R\,w-\frac{1}{2}w^{\ast }KRK^{\ast }w\right) \,\frac{d^{2s}z}{\pi ^{s}}
\notag \\
&=&c^{2}\int \exp \left( i\,\mathrm{Im}{z^{\ast }}(I_{s}+K^{\ast })w-\left(
z+\frac{I_{s}+K^{\ast }}{2}w\right) ^{\ast }Rz+\frac{I_{s}+K^{\ast }}{2}%
w\right)  \notag \\
&&\times \exp \left( -w^{\ast }\left( \frac{I_{s}-K}{2}\right) R\left( \frac{%
I_{s}-K^{\ast }}{2}\right) w\right) \,\frac{d^{2s}z}{\pi ^{s}}  \notag \\
&=&\exp \left( -w^{\ast }Bw\right) ,  \label{bg}
\end{eqnarray}%
which is Gaussian characteristic function with
\begin{equation}
B=\frac{1}{4}\left[ \left( I_{s}+K\right) R^{-1}\left( I_{s}+K^{\ast
}\right) +\left( I_{s}-K\right) R\left( I_{s}-K^{\ast }\right) \right] .
\label{be}
\end{equation}%
Then (\ref{gd}) with (\ref{bg}) mean that $\Phi _{M}$ is a gauge-covariant
Gaussian channel. $\square $

The matrix (\ref{be}) of the quadratic form in the exponent (\ref{bg}) must
satisfy the general necessary and sufficient condition for quantum channels
(see Eq. (12.170) in \cite{QSCI})
\begin{equation}
B \geq \pm \frac{1}{2}\left( I_{s}-KK^{\ast }\right) .  \label{nsc}
\end{equation}%
Although this should follow automatically from the complete positivity of
the map (\ref{M-add}), let us give an independent check. Denote by $r_{j}$
the eigenvalues, $v_{j}$ the eigenvectors of the positive definite Hermitian
matrix $R=2L+I_{s}.$ Let $z$ be an arbitrary vector from $\mathbb{C}^{s}$
and $w=K^{\ast }z.$ Then (\ref{nsc}) amounts to%
\begin{equation*}
\left( z+w\right) ^{\ast }R^{-1}\left( z+w\right) +\left( z-w\right) ^{\ast
}R\left( z-w\right) \geq \pm 2\left( z^{\ast }z-w^{\ast }w\right) ,
\end{equation*}%
or
\begin{equation*}
\sum_{j=1}^{s}\left[ r_{j}^{-1}\left\vert a_{j}+b_{j}\right\vert
^{2}+r_{j}\left\vert a_{j}-b_{j}\right\vert ^{2}\right] \geq \pm
2\sum_{j=1}^{s}\left( \left\vert a_{j}\right\vert ^{2}-\left\vert
b_{j}\right\vert ^{2}\right) ,
\end{equation*}%
where $a_{j}=v_{j}^{\ast }z,\,b_{j}=v_{j}^{\ast }w.$ But for arbitrary $r>0$
and complex $a,\,b$
\begin{equation*}
r^{-1}\left\vert a+b\right\vert ^{2}+r\left\vert a-b\right\vert ^{2}\geq
2\left\vert \left\vert a\right\vert ^{2}-\left\vert b\right\vert
^{2}\right\vert ,
\end{equation*}%
implying the required inequality.

\section{One mode}

\label{s5}

\begin{figure}[t]
\center{\includegraphics{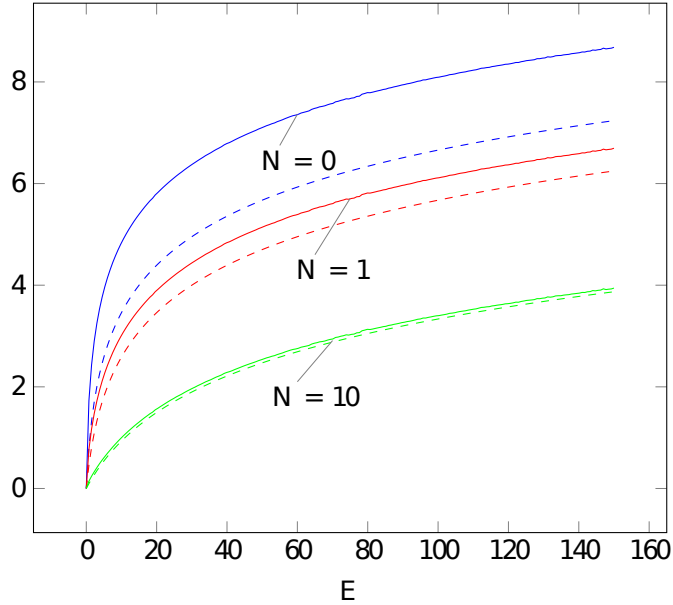}}
\caption{(color online) Comparison of the assisted capacity $C_{ea}$ (solid
line) and unassisted capacity $C$ (dotted line).}
\label{fig0}
\end{figure}

\begin{figure}[t]
\center{\includegraphics{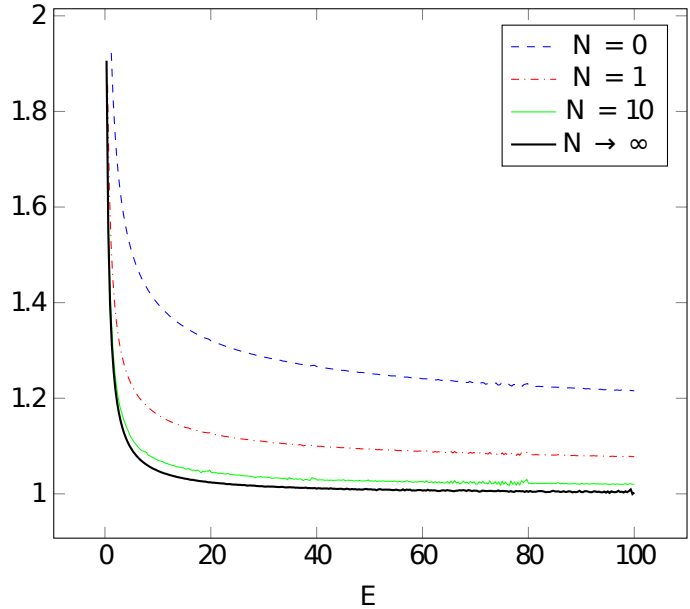}}
\caption{(color online) The gain of entanglement assistance $G=C_{ea}/C$.}
\label{fig1}
\end{figure}

In this section we consider the channel $\tilde{\mathcal{M}}$ defined by the
POVM (\ref{gauss_POVM}) with $s=1$. Take the energy operator $H=a^{\dagger
}a,$ then theorem \ref{t1} and theorem \ref{t2} imply that the maximum in (%
\ref{cmf}) is attained on the gauge-invariant Gaussian state $\rho _{\Lambda
}$ with $\Lambda =E$ (see remark just before theorem \ref{t2}), and the
energy-constrained entanglement-assisted capacity is given by the formula
\begin{equation}
C_{ea}\equiv C_{ea}(\tilde{\mathcal{M}},H,E)=g(E)-g\left( \frac{NE}{N+E+1}%
\right) .  \label{cea_M}
\end{equation}%
For $N=0$ we recover the formula $C_{ea}=g(E)$ obtained in \cite{h5}.

Let us compare (\ref{cea_M}) with the unassisted capacity, which was
calculated in \cite{acc}, i.e.
\begin{equation*}
C\equiv C(\tilde{\mathcal{M}},H,E)=\log \left( N+E+1\right) -\log \left(
N+1\right) .
\end{equation*}%
We will use the asymptotic
\begin{equation*}
g(E)=(E+1)\log (E+1)-E\log E\,\sim \left\{
\begin{array}{c}
-E\log E,\quad E\rightarrow 0 \\
\log E,\quad E\rightarrow \infty%
\end{array}%
\right. .
\end{equation*}

Then it is easy to see that in the limit $E\rightarrow 0$ (weak signal,
noise $N$ fixed)
\begin{equation*}
C\sim \frac{E}{N+1}\log e,\quad C_{ea}\sim -\frac{E}{N+1}\log E,
\end{equation*}%
so that the gain of entanglement assistance
\begin{equation*}
G=C_{ea}/C\sim -\ln E,~E\rightarrow 0.
\end{equation*}

When $E\rightarrow \infty $ (strong signal) we have
\begin{equation}
C_{ea}\sim C\sim \log E,  \label{asymp_cea1}
\end{equation}%
so that $G\sim 1,$ while
\begin{equation*}
C_{ea}-C=\log \left( 1+\frac{1}{E}\right) ^{E}-\log \left( 1+\frac{N+E+1}{NE}%
\right) ^{\frac{NE}{N+E+1}},
\end{equation*}%
and
\begin{equation*}
\lim_{E\rightarrow \infty }(C_{ea}-C)=\log e-\log \left( 1+\frac{1}{N}%
\right) ^{N}.
\end{equation*}

Another interesting limit is $N\rightarrow \infty $ (weak noise, $E$ fixed).
Using the relation $g^{\prime }(E)=\log \left( \frac{E+1}{E}\right) ,$ we
obtain%
\begin{equation*}
C_{ea}\sim g^{\prime }(E)\left( E-\frac{NE}{N+E+1}\right) =\log \left( \frac{%
E+1}{E}\right) \frac{E(E+1)}{N+E+1}
\end{equation*}%
while $C \sim \log e\,\frac{E}{N+1}$ whence
\begin{equation*}
G\sim (E+1)\,\ln \left( \frac{E+1}{E}\right) ,
\end{equation*}%
which varies from $\infty $ for $E\rightarrow 0$ to 1 for $E\rightarrow
\infty .$

The plots of the two capacities and of the gain for the values of
measurement noise $N=0,1,10$ are shown on Fig. \ref{fig0}, \ref{fig1}.

\section{Arbitrary Gaussian measurements}

\label{s6}

Consider $2s$-dimensional symplectic space $(\mathbb{R}^{2s},\Delta )$ with
\begin{equation}
\Delta =\mathrm{diag}\left[
\begin{array}{cc}
0 & 1 \\
-1 & 0%
\end{array}%
\right] _{j=1,\dots ,s}.  \label{delta}
\end{equation}%
In order to spare the symbols, we will preserve the notations for the real
vector $z=[x_{j},y_{j}]_{j=1,\dots ,s}^{t}\in \mathbb{R}^{2s}$ and for the
corresponding element of the volume. Let $\mathcal{H}$ be the space of an
irreducible representation $z\rightarrow W(z);\,z\in \mathbb{R}^{2s},$ of
the Weyl-Segal canonical commutation relations
\begin{equation}
W(z)W(z^{\prime })=\exp [-\frac{i}{2}z^{t}\Delta z^{\prime }]\,W(z+z^{\prime
}).  \label{weyl}
\end{equation}%
Here $W(z)=\exp i\,\sum_{j=1}^{s}(x_{j}q_{j}+y_{j}p_{j})$ are the unitary
Weyl operators, where $q_{j},\,p_{j}$ are the canonical observables of the
quantum system. We denote by $\rho _{\alpha },\,\rho _{\beta }$ centered
Gaussian states with correlation matrices $\alpha ,\beta $ (see e.g. ch.12
of \cite{QSCI} for a detailed description).

We will consider the Gaussian observable given by the POVM
\begin{equation}
M(d^{2s}z)=W(z)\rho _{\beta }W(z)^{\ast }\frac{d^{2s}z}{\left( 2\pi \right)
^{s}}  \label{MTB1}
\end{equation}%
and the corresponding measurement channel $\mathcal{M}$ (see e.g. \cite%
{acc-noJ}). Let $\mathfrak{S}(\alpha )$ be the set of all of centered states
$\rho $ with correlation matrix $\alpha .$ We will study the following
entropic characteristic of the Gaussian measurement channel $\mathcal{M}$
underlying its entanglement-assisted capacity
\begin{equation}
ER(\mathcal{M};\alpha )=\sup_{\rho \in \mathfrak{S}(\alpha )}ER(\rho ,%
\mathcal{M}).  \label{cchi1}
\end{equation}

\begin{theorem}
\label{t3} \emph{The supremum in (\ref{cchi1}) } \emph{is attained on the
Gaussian state $\rho _{\alpha }$ and is equal to
\begin{equation}
ER(\mathcal{M};\alpha )=\frac{1}{2}\left[ \mathrm{Sp\,}g\left( \mathrm{abs}%
(\Delta ^{-1}\alpha )-\frac{I_{2s}}{2}\right) -\mathrm{Sp\,}g\left( \mathrm{%
abs}(\Delta ^{-1}\tilde{\alpha})-\frac{I_{2s}}{2}\right) \right]
\label{cchi2}
\end{equation}%
where
\begin{equation}
\tilde{\alpha}=\beta -\sqrt{I_{2s}+\left( 2\beta \Delta ^{-1}\right) ^{-2}}%
\beta \left( \alpha +\beta \right) ^{-1}\beta \sqrt{I_{2s}+\left( 2\Delta
^{-1}\beta \right) ^{-2}}  \label{121}
\end{equation}%
and }$\mathrm{abs}(\Delta ^{-1}\alpha )$ \emph{is the matrix with
eigenvalues equal to moduli of eigenvalues of } $\Delta ^{-1}\alpha $ \emph{\
and with the same eigenvectors.}
\end{theorem}

In this theorem we do not assume the gauge symmetry: $\alpha $ and $\beta $
need not share the common complex structure. Notice that in the
gauge-invariant case we have the correspondence $\alpha \rightarrow \Lambda
+I_{s}/2$, $\beta \rightarrow N+I_{s}/2,$ $\,\Delta ^{-1}\beta \rightarrow
i\left( N+I_{s}/2\right) $ \cite{QSCI}, and (\ref{121}) turns into (\ref%
{tildeN}).

\emph{Proof (sketch).} For the difference $ER(\rho _{\alpha },\mathcal{M}%
)-ER(\rho ,\mathcal{M})$ we have a representation similar to (\ref{difer}).
It follows that to prove $ER(\rho _{\alpha },\mathcal{M})-ER(\rho ,\mathcal{M%
})\geq 0,\,\rho \in \mathfrak{S}(\alpha ),$ it is sufficient to establish
the analog of (\ref{res}) i.e.
\begin{equation}
\int \mathrm{Tr}(\rho _{\alpha }(z)-\rho (z))\log \,\hat{\rho}_{\alpha }(z)\,%
\frac{d^{2s}z}{\left( 2\pi \right) ^{s}}=0,  \label{res1}
\end{equation}%
where $\rho (z)=\sqrt{\rho _{\beta }}W(z)^{\ast }\rho W(z)\sqrt{\rho _{\beta
}},\,\rho _{\alpha }(z)=\sqrt{\rho _{\beta }}W(z)^{\ast }\rho _{\alpha }W(z)%
\sqrt{\rho _{\beta }},\,\hat{\rho}_{\alpha }(z)=\rho _{\alpha }(z)/p_{\alpha
}(z)$ and $p_{\alpha }(z)=\mathrm{Tr}\rho _{\alpha }(z)=\mathrm{Tr}\rho
_{\alpha }W(z)\rho _{\beta }W(z)^{\ast }.$

To establish (\ref{res1}), we first prove generalization of lemma \ref{l1}:
for the input state $\rho =\rho _{\alpha },$ the posterior states are
Gaussian, namely%
\begin{equation*}
\hat{\rho}_{\alpha }(z)=W(Kz)^{\ast }\rho _{\tilde{\alpha}}W(Kz),
\end{equation*}%
where $K$ is a real square matrix and $\tilde{\alpha}$ is real correlation
matrix (\ref{121}) of the centered Gaussian state $\rho _{\tilde{\alpha}}.$
This is established with the help of the formula for the characteristic
function of product of Gaussian states established in the Appendix of \cite%
{hsh}. More specifically, the correlation matrix of the operator $\sqrt{\rho
_{1}}\rho _{2}\sqrt{\rho _{1}}$ where $\rho _{1},\rho _{2}$ are Gaussian,
was computed in \cite{scut} (see also \cite{pira}) and the formula (\ref{121}%
) was given in \cite{lami}, eq. (3.27).

Then similarly to (\ref{res3}) we have
\begin{equation}
\int \mathrm{Tr}(\rho _{\alpha }(z)-\rho (z))\log \,\hat{\rho}_{\alpha }(z)\,%
\frac{d^{2s}z}{\left( 2\pi \right) ^{s}}=\mathrm{Tr}\,\left[ \Phi
_{M}\,(\rho _{\alpha })-\Phi _{M}(\rho )\right] \log \rho _{\tilde{\alpha}},
\label{res4}
\end{equation}%
where
\begin{equation*}
\Phi _{M}(\sigma )=\int W(Kz)\sqrt{\rho _{\beta }}\,W(z)^{\ast }\,\sigma
\,W(z)\sqrt{\rho _{\beta }}\,W(Kz)^{\ast }\frac{d^{2s}z}{\left( 2\pi \right)
^{s}}.
\end{equation*}%
The proof that $\Phi _{M}$ is Gaussian channel and hence the right-hand side
of (\ref{res4})\ is equal to zero follows the same lines as in lemma \ref{l2}.

This proves
\begin{eqnarray*}
ER(\mathcal{M};\alpha ) &=&ER(\rho _{\alpha },\mathcal{M}) \\
&=&\mathrm{H}(\rho _{\alpha })-\int p_{\alpha }(z)\mathrm{H}(W(Kz)^{\ast
}\rho _{\tilde{\alpha}}W(Kz))\frac{d^{2s}z}{\left( 2\pi \right) ^{s}} \\
&=&\mathrm{H}(\rho _{\alpha })-\mathrm{H}(\rho _{\tilde{\alpha}}).
\end{eqnarray*}

The formula (\ref{cchi2}) now follows from the expression for the entropy of
an arbitrary Gaussian state given by eq. (12.110) in  \cite{QSCI}. $\square $

\textbf{Acknowledgment}. The work was supported by the grant of Russian
Science Foundation (project No 19-11-00086). The authors are grateful to
M.E. Shirokov for useful comments.

\end{document}